\documentclass[final]{aipproc}
\layoutstyle{6x9}
\usepackage{amssymb}

\newcommand{\apj}{Astrophysical Journal}

\newcommand{\nat}{Nature}
\newcommand{\aaa}{Astronomy \& Astrophysics}

\begin{document}

\title{Accretion in the spin-down regime}

\bibliographystyle{aipproc} 

%\author{Ek\c{s}i, Kutlu}


\author{Ek\c{s}i, K.~Y.}{
  address={\.Istanbul Technical University, Faculty of Arts and Letters, Department of Physics}
}

\author{Kutlu, E.}{
  address={\.Istanbul Technical University, Faculty of Arts and Letters, Department of Physics}
}

%\begin{document}
%\maketitle

\begin{abstract}

Accretion driven millisecond X-ray pulsars can accrete over a wide range of mass flow rates.
The pulsations persist even at small accretion rates at which these objects would be 
expected to be in the propeller stage.
We argue that the inner regions of disks around millisecond X-ray pulsars are sufficiently 
thick that a fraction of the inflowing matter can accrete even in the spin-down regime 
from regions of the disk away from the disk plane. 
This allows these systems to be bright throughout a wide range of mass flow rates. 
We model the lightcurve of SAX J1808.4-3658 during an outburst and show that the rapid decay stage can be modeled with 
fractional accretion in the spin-down regime.
 
\end{abstract}

\maketitle

\section{Introduction}

The discovery of accretion-powered millisecond
X-ray pulsars (AMXP) \cite{int98,wij98} supports the recycling scenario \cite{alp82,rad82} 
according to which
millisecond radio pulsars \cite{bac82} are descendants of 
low mass X-ray binaries.

All of the 13 AMXPs known to date are X-ray transients. During an outburst the luminosity of these objects 
can change two orders of magnitude. According to the conventional picture classifying
neutron stars in terms of their interaction with the surrounding matter, as the mass flow rate
$\dot{M}$ declines throughout the outburst, these objects
should undergo a transition to the propeller stage 
\cite{ill75} in which the centrifugal barrier inhibits inflowing matter in the
disk to reach the surface of the neutron star.
Around this transition the torque should also change sign so that the star spins down. 

Accretion onto AMXPs even at low accretion rates is a challenge to this picture.
In some cases, even while the  
the neutron stars exhibit a high rate of spin-down, accretion onto the neutron star continues, and
the X-ray pulsations prevail.
In order to explain these difficulties Rappaport et al.~\cite{rap04} has presented a 
steady state solution of the disk structure in which the inner radius of the disk remains 
on the co-rotation radius while $\dot{M}$ declines. 

SAX J1808.4-3658 experiences outbursts
lasting a few weeks roughly once in two years, during
which the coherent $\sim 401$ Hz pulsations are observable.
The inner radius of the accretion disk,
$R_{\mathrm{in}} = \xi (\mu^2/\sqrt{2GM_{\ast}}\dot{M})^{2/7}$
where $\mu$ is the magnetic moment, $M_{\ast}$ is the mass of the neutron star 
and $\xi$ is a factor of unity, is expected to increase with
dropping $\dot{M}$. 
There is recent evidence \cite{ibr09,strat} that
the $R_{\mathrm{in}}$ \emph{does} recede during the decline of the outburst.

An \emph{infinitely thin} disk can not accrete 
matter as soon as $R_{\mathrm{in}}$ moves beyond the co-rotation radius
$R_{\mathrm c} =(GM_{\ast}/\Omega_{\ast}^2)^{1/3}$
where  $\Omega_{\ast}$ is the angular frequency of the neutron star.
In this case the fastness parameter
$\omega_{\ast}  = (R_{\mathrm{in}}/R_{\rm c})^{3/2}$
becomes greater than unity. The luminosity is expected to drop to zero abruptly.
The situation can be different for a disk with a \emph{finite} thickness which will be able to accrete matter 
even for $\omega_{\ast} \gtrsim 1$ because the
part of the disk at high latitudes will not be centrifugally inhibited. 
Inclination between the rotation axis and the magnetic axis will further facilitate accretion.

Here, we reassess the work of \cite{men99} for estimating the fraction of mass 
that can accrete in the spin down regime and apply it to model the lightcurve of  SAX J1808.4-3658.

\section{Matter accreting in the spin-down regime}

The fraction of matter that can accrete onto a neutron star in the spin-down regime 
was first estimated by Lipunov \& Shakura \cite{lip76} for spherical accretion. 
Here we follow the work of Menou et al.~\cite{men99}
for quasi-spherical disk accretion \cite{nar95}. There are recent numerical studies \cite{rom04,ust06}
that address accretion in the spin-down regime.

The magnetic surface is defined as
$r_{\mathrm{in}} = R_{\mathrm{in}}(1+3\cos^2 \theta)^{2/7}$
where $\theta$ measures the angle from the rotation axis which we assume to coincide with the magnetic axis.  
The co-rotation surface is defined as
$r_{\mathrm c} = R_{\mathrm c} \sin^{-2/3} \theta$.
A disk with a \emph{finite} thickness will be able to accrete matter even 
when $R_{\mathrm{in}}>R_{\mathrm c}$ because at high latitudes ($0<\theta<\theta_0$)
$r_{\mathrm{in}}<r_{\mathrm c}$ can be satisfied.
The fraction of the inflowing matter that can accrete onto the neutron star is
\begin{equation}
f \equiv \frac{\dot{M}_{\ast }}{\dot{M}}=\frac{2\int_{0}^{\theta _{0}} 
\rho v 2\pi r^{2}\sin \theta d\theta }{2\int_{0}^{\pi /2} \rho v 2\pi r^{2}\sin \theta d\theta }.
\end{equation}
Here the critical angle $\theta_0$ is defined through $r_{\mathrm A} = r_{\rm c}$. 
Given the uncertainty of the size of the magnetosphere in the presence of plasma, 
we follow \cite{men99} and assume that $r_{\mathrm{in}}\simeq R_{\mathrm{in}}$ which gives 
$\sin \theta_0 = \omega_{\ast}^{-1}$. Using $\rho (r,\theta )\simeq \rho (r)$ and 
$v(r,\theta )\simeq v(r)\sin ^{2}\theta$ from the quasi-spherical solutions of \cite{nar95}, one obtains
\begin{equation}
f = 1-\frac{3}{2}\left( 1-\omega_{\ast }^{-2}\right)^{1/2}+\frac{1}{2}
\left( 1-\omega _{\ast}^{-2}\right)^{3/2} 
\label{frac}
\end{equation}
Note that this result follows from the work of \cite{men99}, the only difference being
that these authors, at the final step, used the small angle approximation 
(i.e. $\theta_0 \ll 1$) and so found $f =(3/8) \omega_{\ast}^{-4}$  
which is $\omega_{\ast} \gg 1$ limit of the above.

%%%%%%%%%%%%%%%%%%%%%%%%%%%%%%%%%%
%%%%%%%%%%%%%%%%%%%%%%%%%%%%%%%%%%
\section{Evolution of the disk}
%%%%%%%%%%%%%%%%%%%%%%%%%%%%%%%%%%
%%%%%%%%%%%%%%%%%%%%%%%%%%%%%%%%%%
The thick inner region of the disk will match to a standard thin disk \cite{sha73} outside.
The evolution of $\dot{M}$ during the outburst will be determined by what the outer thin disk provides
to the inner quasi-spherical region.
The evolution of a thin disk is described by the diffusion equation
\begin{equation}
\frac{\partial \Sigma }{\partial t}=\frac{3}{r}\frac{\partial }{\partial r}
\left[ r^{1/2}\frac{\partial }{\partial r}\left( r^{1/2}\nu \Sigma \right) \right]
\label{diffusion} 
\end{equation}
where $\Sigma$ is the surface mass density and $\nu$ is the turbulent viscosity. 
For a standard thin disk \cite{sha73}, this equation has three self-similar 
solutions \cite{pri74} two of which have freely expanding outer radius and are 
not suitable for tidally truncated disks in binaries. The third solution is suitable for
 disks in binary systems \cite{lip00}. According to this solution
\begin{equation}
\dot{M} = \dot{M}_0 \left( 1 + \frac{t}{t_0}  \right)^{-\alpha}
\label{inflow}
\end{equation} 
where $\dot{M}_0$ is the mass flow rate at the beginning of the outburst, $t_0$ is 
the viscous time-scale in the outer disk, and $\alpha$ depends on the opacity and 
pressure regime prevailing in the disk.
For an electron scattering and gas pressure dominated disk $\alpha=5/2$. 

The X-ray luminosity of accretion onto the neutron star is found by
$L_{\rm X} = GM \dot{M}_{\ast}/R$
where $\dot{M}_{\ast}=\dot{M}$ if $\omega_{\ast}<1$ and $\dot{M}_{\ast}=f\dot{M}$ if $\omega _{\ast }>1$.

%%%%%%%%%%%%%%%%%%%%%%%%%%%%%%%%%%%%%%%%%%%
%%%%%%%%%%%%%%%%%%%%%%%%%%%%%%%%%%%%%%%%%%%
\section{Application to SAX J1808.4-3658}
%%%%%%%%%%%%%%%%%%%%%%%%%%%%%%%%%%%%%%%%%%%
%%%%%%%%%%%%%%%%%%%%%%%%%%%%%%%%%%%%%%%%%%

\begin{figure}[t]
  \includegraphics[height=.5\textwidth]{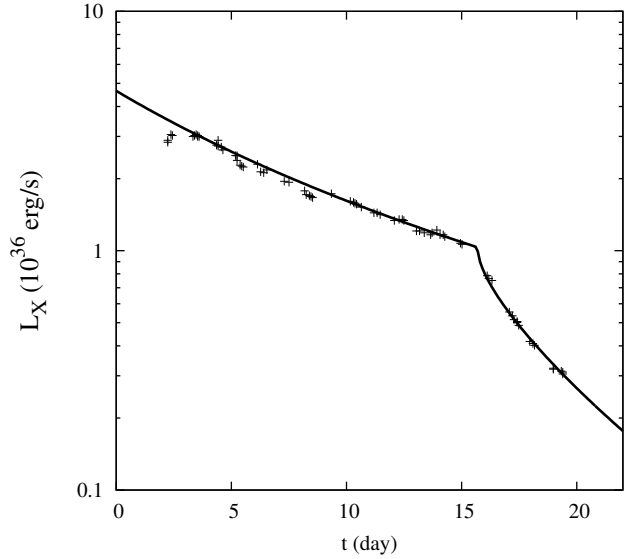}
  \caption{The lightcurve of SAX J1808.4--3658 in October--November 2002 outburst compared with the model presented in this work. 
The data is taken from the first figure in \cite{ibr09} excluding the flaring decay stage which is not addressed here. The peak stage can not be modeled with the self-similar model for $\dot{M}$ employed in this work. The slow decay stage is modeled by assuming all the inflowing mass accretes onto the star. As $\dot{M}$ declines $\omega_{\ast}$ grows beyond unity. In the rapid decay stage only a fraction of $\dot{M}$ can accrete.}
\end{figure}

SAX J1808.4--3658 showed an outburst in October--November 2002 followed by Rossi X-ray Timing Explorer.
The lightcurve shows 4 stages: peak, slow decay, rapid decay, and flaring decay as described by Ibragimov \& Poutanen~\cite{ibr09}. 
We converted the $3-20$ keV flux data therein to luminosity by assuming the distance to be 
3.5 kpc \cite{gal06}.

The slow decay region can be modeled by
using $t_0=19$ days, $\dot{M}_0=2.2\times 10^{16}$ g s$^{-1}$ and $\alpha=5/2$ in Eqn.(\ref{inflow})
assuming all matter can acccrete.

We associate the commence of the rapid decay with $\omega_{\ast}=1$ which allows for determining 
the magnetic moment of the star as $\mu = 0.7\xi^{-7/4}\times 10^{26}$ G cm$^3$.
For $\xi=0.5$ 
this corresponds to a magnetic moment of $\mu = 2.2\times 10^{26}$ G cm$^3$.
We assume that $\dot{M}$ declines at the same rate given in Eqn.~(\ref{inflow}), 
but now a fraction, determined by Eqn.(\ref{frac}), of it accretes onto the neutron star.
We find that the rapid decay stage can be modeled as fractional 
accretion in the fast ($\omega_{\ast}>1$) regime.

Because of the self-similar model employed for $\dot{M}$, 
the rise to the peak stage cannot be addressed in this work. This can be done by 
solving Eqn.~(\ref{diffusion}) numerically with an appropriate 
initial distribution for $\Sigma$. The rapid decay stage is followed 
by a flaring decay stage (see the Figure 1 in \cite{ibr09}) which is not addressed in this work. 

\section{Discussion}

We have suggested that the rapid decay stage of the lightcurve of SAX J1808.4--3658 during an outburst
can be modeled by fractional accretion in the rapidly rotating stage ($\omega_{\ast}>1$). 
This requires the inner region of the disk to be sufficiently thick.  
Our result, $B=1.4\xi^{-7/4}\times 10^8$ G, is consistent
with earlier estimates \cite{gil98,psa99,har08}.

%%%%%%%%%%%%%%%%%%%%%%%%%%%%%%%%%%%%%%%%%%%%%%%%
%% BACKMATTER
%%%%%%%%%%%%%%%%%%%%%%%%%%%%%%%%%%%%%%%%%%%%%%%%

\begin{theacknowledgments}
We thank M.~Ali Alpar for his suggestions in improving the manuscript. 
KYE acknowledges support from the Faculty of Arts and Letters in \.IT\"U and
Marie Curie EC FPG Marie Curie Transfer of Knowledge Project 
ASTRONS, MKTD-CT-2006-042722.
\end{theacknowledgments}


\begin{thebibliography}{99}

\bibitem{int98}
J.~J.~M. in't Zand et al.~, \emph{\aaa}, \textbf{331}, 25--28 (1998).

\bibitem{wij98}
R.~Wijnands, and M.~van der Klis, \emph{\nat}, \textbf{394}, 344--346 (1998).

\bibitem{alp82}
M.~A. Alpar, A.~F. Cheng, M.~A. Ruderman, and J.~Shaham,
\emph{\nat}, \textbf{300}, 728--730 (1982).

\bibitem{rad82}
V.~Radhakrishnan, and G.~Srinivasan, \emph{Current Science}, \textbf{51}, 1096--1099 (1982).

\bibitem{bac82}
D.~C. Backer, S.~R. Kulkarni, C.~Heiles, M.~M. Davis, and W.~M. Goss,
\emph{\nat}, \textbf{300}, 615--618 (1982).

\bibitem{ill75}
A.~F. Illarionov, and R.~Sunyaev, \emph{\aaa}, \textbf{39}, 185--195 (1975).

\bibitem{rap04}
S.~A. Rappaport, J.~M. Fregeau, and H.~Spruit, \emph{\apj}, \textbf{606}, 436--443 (2004).

\bibitem{ibr09}
A.~Ibragimov, and J.~Poutanen, \emph{MNRAS},\textbf{400}, 492--508 (2009).

\bibitem{strat}
S.~van Straaten,  M.~van der Klis, and R.~Wijnands, \emph{\apj}, \textbf{619}, 455--482 (2005).

\bibitem{men99}
K.~Menou et al.~, \emph{\apj}, \textbf{520}, 276--291 (1999).

\bibitem{lip76}
V.~M. Lipunov, and N.~I. Shakura, 
\emph{Soviet Astronomy Letters}, \textbf{2}, p. 133--135 (1976).

\bibitem{rom04}
M.~M. Romanova et al.~, 
\emph{\apj}, \textbf{616}, L151--L154 (2004).

\bibitem{ust06}
G.~V. Ustyugova et al.~,
\emph{\apj}, \textbf{646}, 304--318 (2006).


\bibitem{nar95} 
R.~Narayan, and I.~Yi, \emph{\apj}, \textbf{444}, 231--243 (1995).

% Evolution of the disk

\bibitem{sha73}
N.~I. Shakura, and R.~Sunyaev, \emph{\aaa}, \textbf{24}, 337--355 (1973).
 

\bibitem{pri74}
J.~E. Pringle, \emph{Ph.D. thesis}, Univ. Cambridge, (1974).

\bibitem{lip00}
G.~V. Lipunova, and N.~I. Shakura, \emph{\aaa}, \textbf{356}, 363--372 (2000).


% Application
\bibitem{gal06}
D.~K. Galloway, A.~Cumming, \emph{\apj}, \textbf{652}, 559--568 (2006).

\bibitem{gil98}
M.~Gilfanov et al.~, 
 \emph{\aaa}, \textbf{338}, 83--86 (1998).
 
 \bibitem{psa99} 
D.~Psaltis, and D.~Chakrabarty, \emph{\apj}, \textbf{521}, 332--340 (1999).

\bibitem{har08}
 J.~M. Hartman  et al.~, \emph{\apj}, \textbf{675}, 1468--1486 (2008).


\end{thebibliography}
\end{document}